\documentclass[preprint,showpacs,titlepage,preprintnumbers,aps,prd,amsmath,amssymb,byrevtex,nofootinbib]{revtex4}
\bibpunct{[}{]}{,}{n}{}{;}
\usepackage{graphicx}
\usepackage{natbib}
\usepackage{bm}

\begin{document}

\preprint{FERMILAB-PUB-05-048-CD}\preprint{physics/0503229}
\vspace*{-2cm}

\title{\Large The Improbability Scale}
\author{\large David J. Ritchie}
\vspace*{1.0cm}
\affiliation{\sl Computing Division\\
Fermi National Accelerator Laboratory\\
P.~O.~Box~500, Batavia, IL 60510, USA\\
ritchie@fnal.gov}

\date{March 30, 2005\\[1cm]} 

\begin{abstract}
The Improbability Scale ($IS$) is proposed as a way of communicating to the general public the improbability (and by implication, the probability) of events predicted as the result of scientific research. Through the use of the Improbability Scale, the public will be able to evaluate more easily the relative risks of predicted events and draw proper conclusions when asked to support governmental and public policy decisions arising from that research.
\end{abstract}

\vglue 1.4cm
\maketitle


\section{Introduction}
Many aspects of life as a responsible citizen in society
involve having an understanding of the probability
of one type of event in comparison to others. Yet event probabilities are 
often expressed using unfamiliar or varied terminology (\textit{i.e.,} 
negative exponents, such as $10^{-4}$ or $10^{-5}$,  one part in a thousand, etc.) with the 
result that,
for the ordinary person, the comparison of event probabilities and the drawing of valid conclusions are made more difficult.

\section{Proposed Remedy}
As a remedy for this, I propose the Improbability Scale, or $IS$, defined as: 
{
\mathversion{bold}
\begin{equation}
IS = - \log_{10} (p) 
\label{mp_master}
\end{equation}
}
where $p$ is the probability of the event. 

$IS$ takes on the value of 0 for absolutely certain events and proceeds upwards for events with greater and greater {\it im}probability.
Table~I lists some events and their $IS$ values.

\begin{center}
\bigskip
\begin{tabular}{|l|c|}
\hline 
\hfil \textbf{Event} \hfil                                                             & \textbf{IS}   \\ \hline
Rolling a 7 on the next roll of a pair of dice~\citep{twodie}                & $0.8$         \\ \hline
Space Shuttle major failure on next launch - current experience~\citep{shuttle}   & $2.3$         \\ \hline
One's birthday occuring tomorrow within a given year~\citep{birthday}        & $2.6$         \\ \hline
Space Shuttle major failure on next launch - near term goal~\citep{shuttle}    & $4.0$         \\ \hline 
Being struck by lightning within a given year~\citep{lightning}            & $5.4$         \\ \hline

Drawing a royal flush on the next deal of five cards~\citep{royalflush}      & $5.8$         \\ \hline
Space shuttle major failure on next launch - eventual goal~\citep{shuttle}      & $6.0$      \\ \hline
Winning the jackpot in the next Powerball Lottery~\citep{powerball}          & $8.1$         \\ \hline
A core-collapse Supernova occurring within a given year close                          &          \\
enough to Earth (8 parsecs) to cause significant biological effects~\citep{supernova}  & $8.8$    \\ \hline
\end{tabular}
\\
\bigskip
Table I \\
Some Events and their $IS$ values\\
\bigskip
\end{center}
Because Improbability Scale values are typically small numbers between $0$ and $10$, they are easily remembered---particularly in the case of personally meaningful events. The public can use the $IS$ values for such  events to ``customize'' its understanding of the Improbability Scale. When a new or less familiar event is presented, the public can use the event{\tt '}s $IS$ value to put its improbability into proper perspective and, by implication, to draw conclusions about the event{\tt '}s {\it probability} as well.

\section{Examples of the Utility of the Improbability Scale}
A standout example of how the Improbability Scale could have served better to communicate the risks of a technological endeavor may be found in an October 2000 speech~\citep{shuttle} on the topic of {\it NASA in the 21st Century} given by then NASA Administrator Daniel Goldin. The speech was given to a Laboratory audience at the Applied Physics Laboratory Colloquium~\cite{APL} of The Johns Hopkins University and was also reported on {\bf Space.com} by Leonard David to a readership more characteristic of the interested general public. There, Goldin is reported as saying: \begin{quote}\begin{center}``We want to take the probability of a major failure of today{\tt '}s space shuttle from one part 
in 200 to one part in 10,000, 
and eventually to one part in 1,000,000 with about the same reliability of today{\tt '}s commercial aircraft.''\end{center}\end{quote} No doubt for the experienced Laboratory audience the implications of a risk assessment of ``one part in 200'' were well understood.  For the interested general public with little or no context in which to place that assessment, the same is not clear.  

However, with context provided by the Improbability Scale and a knowledge of the $IS$ for familiar events, such as that for {\it certainty} equaling $0$ and that for tomorrow being one's birthday equaling $2.6$, the public would have almost certainly understood the implications of a risk assessment that stated: \begin{quote}\begin{center}``On the Improbability Scale, a major failure \\of today{\tt '}s Space Shuttle has a rank of $2.3$.''\end{center}\end{quote}

Another example of the utility of the Improbability Scale relates to the $IS$ for several independent events occuring together. The $IS$ for the combined occurrence is the sum of the $IS$ values for the individual events. This simple combination rule makes it easy for the general public to use its knowledge of the $IS$ values for familiar events to understand the improbability of a new or less familiar event.

Knowing that the $IS$ for one{\tt '}s birthday occuring tomorrow within a given year is $2.6$ and that the $IS$ for being struck by lightning within a given year is $5.4$, one has an immediate understanding of just how improbable an $IS$ $8.0$ event is---namely, it is as improbable as getting struck by lightning on one's birthday. One can then apply that understanding to even mundane matters, such as when one learns that winning the Powerball Lottery jackpot on the next drawing~\citep{powerball} has an $IS$ of $8.1$. 

\section{Conclusion}
I suggest that researchers quote the Improbability Scale values when writing for the general public. Widespread adoption of this way of characterizing events will enhance the public's understanding of the predictions of science and help in obtaining the public's support
for actions related to those predictions in, for example, such cases as natural disasters and technological failures.

\section{Acknowledgements}  
I am grateful to Robert Cousins, Department of Physics and Astronomy, UCLA for a number of discussions. I thank Mariano Zimmler, Division of Engineering and Applied Sciences, Harvard University for a careful reading of the manuscript. Fermilab is operated under DOE contract DE-AC02-76CH03000.


\newpage
\section{References}
\begin{thebibliography}{9}
\bibitem{twodie} 
The probability of rolling two die and having the total 
come out to be seven is $6/36$ for which the $IS$ is computed to be $0.8$. 

\bibitem{shuttle}
Daniel Goldin, as reported by Leonard David, Space.com, October 11, 2000, \url{ http://www.space.com/businesstechnology/business/goldin_tsunami_001011.html}

\bibitem{birthday} 
The probability within the calendar year of today being 
one's birthday is $1 / 365$ or $1 / 366$ depending on whether the year is 
a leap year or not. To two significant figures, the $IS$ is $2.6$.

\bibitem{lightning}
According to the {\it National Lightning Safety Institute}, the number of lightning strike victims in the US per year on average is 1000. Taking the US population at 280,000,000, one finds that the odds of being a victim are 1 in 280,000. From this, one computes the $IS$ to be $5.4$. See: \url{http://www.lightningsafety.com/nlsi_pls/probability.html} 

\bibitem{royalflush} 
The probability of drawing a royal flush is computed 
as $4C1 / 52C5$ or $0.0000015$ from which the $IS$ is computed to be $5.8$.

\bibitem{powerball}
See Powerball Game Information, {\ttfamily http://www.wilottery.com/lottogames/pballinfo.asp}

\bibitem{supernova} 
N.~Gehrels, C.~M.~Laird, C.~H.~Jackman, J.~K.~Cannizzo, B.~J.~Mattson, and W.~Chen,
{}{} ``Ozone Depletion From Nearby Supernovae,'' 
ApJ {\bf 585}, 1169 (2003)
[arXiv:astro-ph/0211361].

\bibitem{APL}
The Johns Hopkins University Applied Physics Laboratory Colloquium began in 1947. Held weekly, it is one of the longest standing technical and scientific lecture series in the Washington/Baltimore area. See:  \url{http://www.jhuapl.edu/colloquium/}

\end{thebibliography}
\end{document}